# Zeptojoule Microwave Calorimetry with Nanosecond Recovery in Graphene

*Seunghan Lee[1], Woochan Jung[1], Jayeon Kim[1], Dongsung T. Park[1], Seong Jang[1,2], Seongeun Jeon[1], Kenji Watanabe[3], Takashi Taniguchi[4], Jong-Chul Park[5], Philip Kim[6], Kin Chung Fong[2,7], and Gil-Ho Lee[1,*]*

[1] Department of Physics, Pohang University of Science and Technology, Pohang 37673, Republic of Korea

[2] Department of Electrical and Computer Engineering, Northeastern University, Massachusetts 02115, USA

[3] Research Center for Electronic and Optical Materials, National Institute for Materials Science, 1-1 Namiki, Tsukuba 305-0044, Japan

[4] Research Center for Materials Nanoarchitectonics, National Institute for Materials Science, 1-1 Namiki, Tsukuba 305-0044, Japan

[5] Department of Physics and Institute for Sciences of the Universe, Chungnam National University, Daejeon 34134, Republic of Korea

[6] Department of Physics, Harvard University, Cambridge, MA 02138, USA

[7] Department of Physics, Northeastern University, Massachusetts 02115, USA

[*]Correspondence to Gil-Ho Lee (lghman@postech.ac.kr)

**Abstract**

Graphene calorimeters are predicted to combine extreme energy sensitivity with rapid thermal recovery, yet experiments remain far from ideal estimates. Here, we directly calibrate a graphene Josephson-junction microwave calorimeter using nanosecond pulses and independently determine the thermal parameters of the same operating device. At 110 mK, calibrated pulse measurements yield a characteristic switching-energy scale of 18.1 meV (2.90 zJ), while time-resolved two-pulse thermometry gives an intrinsic relaxation time of 19.5 ns. Pulse-width-dependent calorimetry and bolometric thermometry show that the electronic heat capacity is 12.4 times and the electron–phonon coupling is approximately 50 times larger than ideal estimates. A thermal model using these measured parameters predicts 21.2 meV, quantitatively reproducing the observed switching response. These results identify the microscopic origin of the sensitivity gap in graphene calorimeters and establish a self-consistent calibration framework for ultrafast nanoscale thermal detectors.

Calorimetric detectors have played a central role in detecting increasingly weak energy-deposition events and uncovering previously inaccessible physical regimes. Their impact spans astrophysics, particle physics, quantum optics, and quantum information processing, where minute amounts of absorbed energy must be converted into measurable signals. Over the past few decades, various detector architectures have steadily reduced the threshold energy ($E_{\mathrm{thres}}$), defined as the minimum detectable deposited energy.

Superconducting detector platforms such as superconducting nanowire single-photon detector [1,2] and microwave kinetic inductance detectors [3,4], have reached $E_{thres}$ ~100 meV, while proximity Josephson junction detectors [5,6] and quantum capacitance detectors [7,8] have pushed $E_{thres}$ down to the few-meV regime. This improvement, however, often comes at the cost of slower recovery, as high responsivity typically requires weak coupling to the dominant relaxation channel, leading to microsecond-to-millisecond dead times in meV-sensitive calorimeters.

Graphene offers a promising platform [9-19] to overcome this trade-off. Its small electronic heat capacity [10,15,17], weak electron–phonon coupling [20-25], and broadband absorption [12] make it attractive for nanoscale calorimetry. Although theoretical studies predict sub-meV sensitivity for graphene Josephson junction (GJJ) calorimeters [12], and previous experiments suggest that $E_{thres}$ ~ 0.1 meV may be achievable under ideal conditions [14,17], calorimetric single-photon detection with GJJs has so far been demonstrated only for near-infrared photons at 1550 nm (~0.8 eV) [18,19]. This large gap between prediction and experiment suggests that the thermal processes governing graphene calorimeters remain incompletely understood. In an ideal calorimeter, an absorbed energy ($E_p$) produces an electron-temperature rise $\Delta T_e \sim E_p/C_e$ and relaxes with thermal relaxation time $\tau_{th}$ (= $C_e/G_{th}$), linking sensitivity directly to the electronic heat capacity $C_e$ and thermal conductance $G_{th}$. Enhanced heat capacity or parasitic dissipation therefore suppresses the temperature response and accelerates relaxation, degrading the detector's performance. However, testing this scaling self-consistently in an operating GJJ has remained difficult because $E_p$, $C_e$, $G_{th}$, and $\tau_{th}$ must all be determined under the same nanosecond-scale measurement conditions.

Here we demonstrate graphene Josephson junction microwave calorimeters and establish a direct pulse-based calibration framework for this platform. Using calibrated microwave pulses, we determine an energy threshold of $E_{thres}$=18.1 meV and a thermal relaxation time of $\tau_{th}$=19.5 ns at 110 mK in a single device. The same measurements allow us to extract the electronic heat capacity and electron–phonon cooling under operating conditions. We find that these thermal parameters are substantially larger than ideal theoretical expectations, explaining the observed sensitivity of the device. Our results establish a self-consistent calibration approach for graphene microwave calorimeters and identify the thermal processes that must be controlled to improve their performance.

## Results and Discussion

### Graphene calorimetry and detection scheme

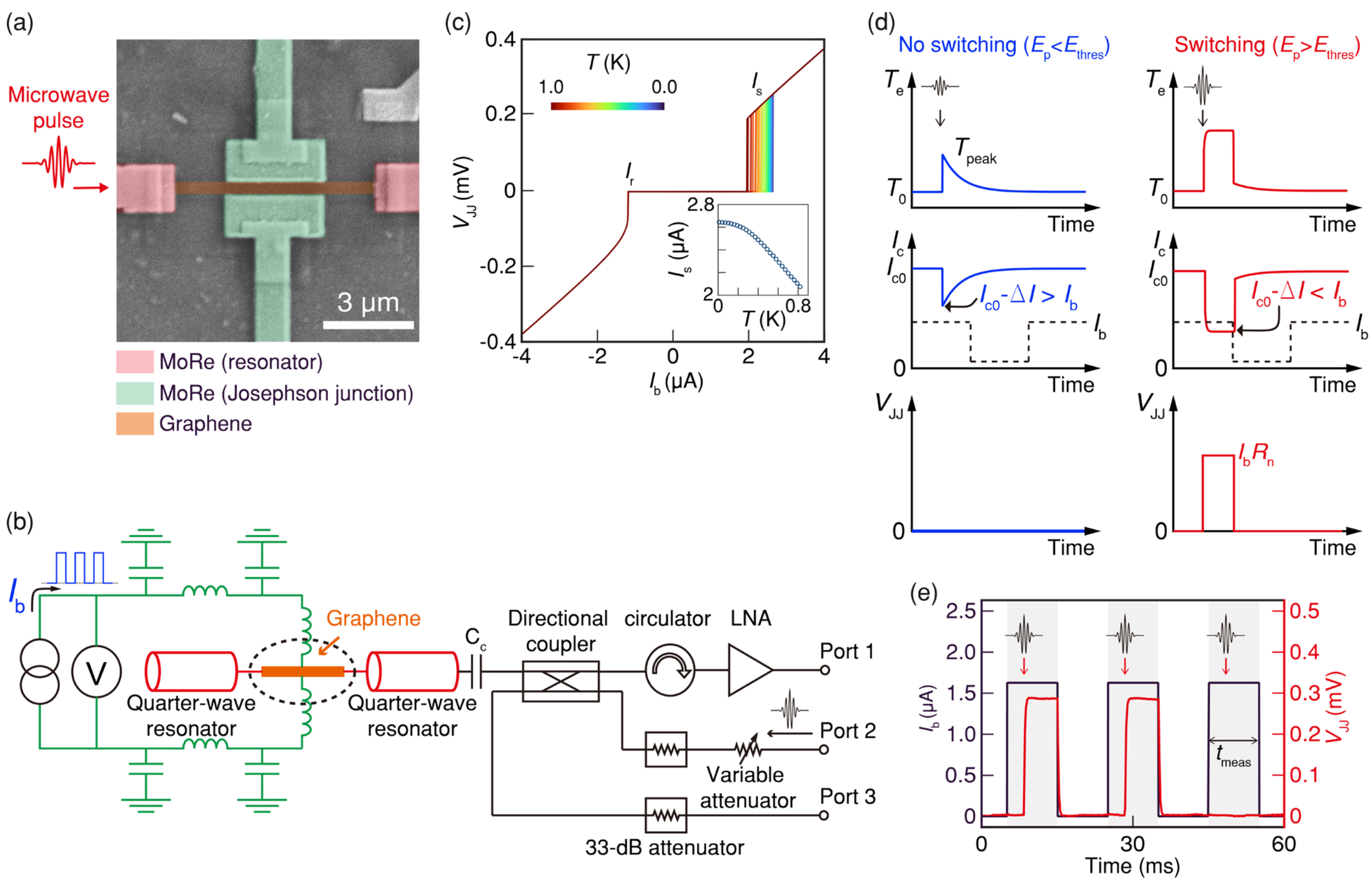

**Figure 1.** Device structure and detection scheme. (a) False-colored scanning electron microscope image of the device, corresponding to the region enclosed by the black dashed circle in (b). (b) Schematic of the measurement setup. Microwave probe tone or pulses are delivered to the detector

through attenuators, a 20-dB directional coupler, and a 200-fF on-chip coupling capacitor ($C_c$), and the reflected signal is amplified by a cryogenic low-noise amplifier (LNA). An on-chip low-pass filter network enables simultaneous d.c. measurement. (c) Current–voltage characteristics of the graphene Josephson junction (GJJ) at temperatures of 0.01-0.82 K. The inset: temperature dependence of the switching current ($I_s$). (d) Detection principle. A pulse increases the electron temperature ($T_e$), reducing the critical current ($I_c$). If the absorbed energy ($E_p$) is insufficient, the junction remains superconducting (left); if $E_p$ exceeds a threshold, the junction switches to the resistive state, producing a finite voltage (right). (e) Example time traces of $I_b$ (black) and $V_{JJ}$ (red) during repeated detection cycles. The shaded region indicates the measurement integration time ($t_{meas}$).

We realized a GJJ integrated with a microwave input resonator in a cross-junction geometry[14], as shown in Figs. 1(a) and (b) ( Methods). The graphene junction is positioned at the current antinode of the resonator to enable efficient absorption of incident microwave signals, while maintaining independent d.c. readout of the junction. This configuration allows controlled energy injection into the graphene electron system and simultaneous monitoring of its electrical response. The detection mechanism relies on the temperature dependence of the critical current ($I_c$) of the GJJ (Fig. 1(c)). The junction is biased with a current ($I_b$) between the critical current at base temperature ($I_{c0}$) and the retrapping current ($I_r$). When energy is deposited in the graphene, the electron temperature ($T_e$) increases, leading to a reduction in $I_c$. If the absorbed energy ($E_p$) is insufficient, the junction remains in the superconducting state and the voltage ($V_{JJ}$) stays zero, while the $T_e$ relaxes back to the base temperature ($T_0$) within the thermal relaxation time. In contrast, if the absorbed energy exceeds a threshold, the junction switches to the resistive state, producing a finite voltage signal (Fig. 1(d)). For hysteretic junctions, the resulting Joule heating further stabilizes the resistive state, enabling a robust voltage readout[23, 26]. The detector is reset by reducing $I_b$ below $I_r$, allowing the junction to return to the superconducting state and prepare for subsequent detection events (Fig. 1(e)).

### Pulse-based direct energy-calibration

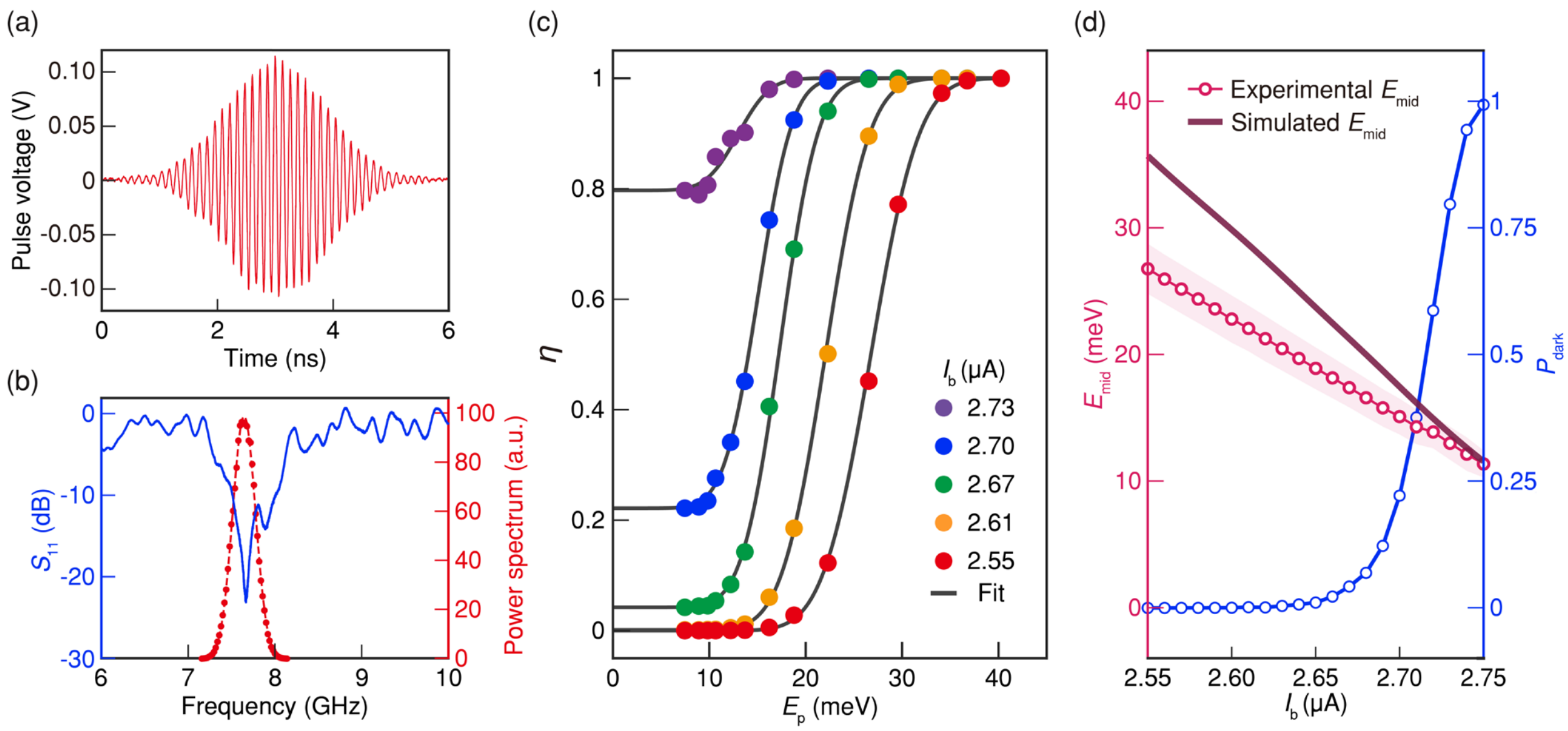

**Figure 2.** Pulse-based calibration of the GJJ detector. (a) Temporal profile of a microwave pulse with a full width at half maximum ($t_p$) of 2 ns. (b) Power spectrum of the pulse (red) and scattering parameter $S_{11}$ of the device (blue), showing a resonance near 7.64 GHz. (c) Detection efficiency ($\eta$) as a function of absorbed pulse energy ($E_p$) for different bias currents ($I_b$). The error bars representing 95% Wilson score confidence intervals are smaller than the symbol size. Solid lines are fits to empirical cumulative distribution functions (CDFs). (d) The median of CDF ($E_{mid}$, left axis) and dark-count probability ($P_{dark}$, right axis) as functions of $I_b$. The shaded region represents the $\pm 0.5\sigma$ range, and the solid line shows the simulation.

We generate Gaussian-enveloped pulses with a center frequency of 7.64 GHz and a full width at half maximum ($t_p$) of 2 ns (Fig. 2(a); see Methods). From independent measurements shown below, we determine $\tau_{th}$ = 19.5 ns, confirming that $t_p \ll \tau_{th}$ under our experimental conditions. The absorbed pulse energy ($E_p$) is calibrated using the same reflectometry-based procedure as in our previous GJJ microwave bolometer[14]. As detailed in Supporting Information (SI) Fig. S1, the resonator response $S_{11}$ is obtained by subtracting the reference trace $S_{13}$, which bypasses the resonator, from the reflected trace $S_{12}$, thereby removing the system background. The resulting $S_{11}$ spectrum is converted into a frequency-dependent absorption spectrum from the reflected power at each frequency. The pulse energy arriving at the device, corrected for the calibrated input-line attenuation, is then weighted by this absorption spectrum over the measured pulse power spectrum to obtain $E_p$. The resonator exhibits a broad 3-dB bandwidth ($\Delta f \sim$ 0.9 GHz; Fig. 2(b)), resulting in a spectrum-averaged absorption of ~ 90 % for the $t_p$ = 2 ns pulses. The detection efficiency ($\eta$), defined as the probability that a pulse induces a switching event, is shown in Fig. 2(c) as a function of $E_p$, for different $I_b$. Each data point is obtained from 10,000 pulse trials. The $\eta$–$E_p$ curves are well described by empirical cumulative distribution functions (CDFs), from which we extract the median of CDF ($E_{mid}$). As $I_b$ approaches $I_{c0}$~3.10 μA, $E_{mid}$ decreases, while the baseline of $\eta$ increases due to an elevated dark-count probability ($P_{dark}$). Figure 2(d) summarizes the dependence of $E_{mid}$ and $P_{dark}$ on $I_b$, highlighting the trade-off between energy sensitivity and dark counts[12]. We define the $E_{thres}$ as $E_{mid}$ at $I_b$ = 2.67 μA, the largest measured $I_b$ for which $P_{dark}$ remains below 5 %, yielding $E_{thres}$ = 18.1 meV. The observed dark counts are attributed to stochastic switching driven by thermal and quantum fluctuations, as well as residual external noise. To quantitatively model the detector response, we numerically solve the thermal equation

$$C_e \frac{dT_e}{dt} = -\Sigma A(T_e^3 - T_0^3) + P_{pulse}(t) \quad (1)$$

where $A$ is the graphene area, $T_0 = 110$ mK is the base electron temperature (see SI Figure S2 for the determination of $T_0$), $P_{\text{pulse}}(t)$ is the absorbed microwave power from Gaussian pulses, temperature-dependent electronic heat capacity is given by $C_e(T_e) = \alpha A \gamma T_e$ with $\gamma = 4\pi^{5/2} k_B^2 n^{1/2}/(3hv_F)$ being the Sommerfeld coefficient—where $k_B$ is the Boltzmann constant, $n$ is the electron density in the graphene channel, $h$ is Planck's constant, and $v_F$ is the Fermi velocity of electrons in graphene— and $\alpha$ a multiplicative scaling factor relative to the theoretical expectation[20], which is extracted from two-pulse measurements discussed below. Here, the cooling is dominated by electron–phonon coupling, as heat diffusion to the superconducting contacts is suppressed by Andreev reflection[10, 23, 27]. Independent bolometric measurements show that a cubic temperature dependence provides a significantly better description of the cooling power than a quartic dependence with $\Sigma = 1.03$ ($\pm 0.02$) W m$^{-2}$ K$^{-3}$ (see SI Fig. S3(f), (g) for the extraction of $\Sigma$). We therefore adopt the $T_e^3 - T_0^3$ form throughout thermal modeling. With these experimentally determined parameters, the temperature rise estimated using Eq. 1 (see SI S4) yields a peak electron temperature ($T_{\text{peak}}$) of ~500 mK for $E_p = 18.1$ meV, which is about 5 % of the superconducting critical temperature of the MoRe contacts ($T_c \approx 10$ K), suggesting that they would continue to serve as effective heat blockers during pulse-induced heating, validating our thermal model[28]. Using these values, the model quantitatively reproduces the measured $E_{\text{mid}}$ (solid line in Fig. 2(d); see SI Fig. S4 for simulation details).

**Pulse-width dependence of detector response**

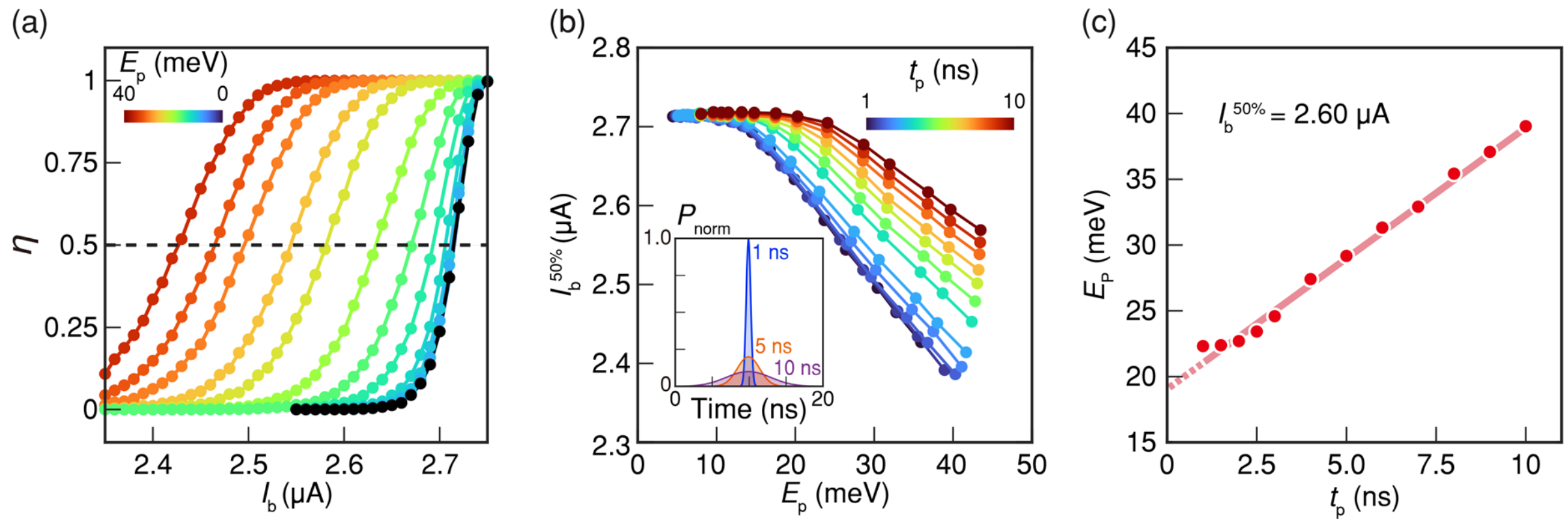

**Figure 3.** Pulse-width dependence of the detector response. (a) Detection efficiency ($\eta$) as a function of bias current ($I_b$) for different absorbed energies ($E_p$) at fixed pulse width ($t_p$ = 1 ns). (b) Bias current corresponding to $\eta$ = 0.5 ($I_b^{50\%}$) as a function of $E_p$ for different $t_p$. The inset: normalized pulse envelopes ($P_{norm}$) for different $t_p$ at constant $E_p$. (c) $E_p$ required to reach $I_b^{50\%}$ = 2.60 μA as a function of $t_p$. The solid line is obtained from the simulation for $t_p \geq 1$ ns, where $t_p$ is much longer than the resonator response time ($\tau_r$). The dashed line is the extrapolation of the simulation.

In our detector scheme, two conditions must be satisfied for a microwave pulse to approximate a single energy-deposition event. First, $t_p$ must be much longer than the resonator response time ($\tau_r$), ensuring that the pulse is delivered to graphene without significant distortion. From the $\Delta f$, we estimate $\tau_r = 1/(2\pi\Delta f) \sim 0.18$ ns, which is well below $t_p$. Second, $t_p$ should be much shorter than $\tau_{th}$. To assess how well this approximation holds in practice, we systematically investigated the dependences of the detector response on $t_p$ and $E_p$. Figure 3(a) shows $\eta$ as a function of $I_b$ for different $E_p$ at fixed $t_p$ = 1 ns. To quantify the horizontal shift of the $\eta$–$I_b$ curves, we define $\boldsymbol{I_b^{50\%}}$ as the bias current corresponding to $\eta$ = 0.5, which serves as a measure of the switching condition for a given $E_p$ [29]. We repeated this analysis for different $t_p$ and extracted the $\boldsymbol{I}_{\mathbf{b}}^{\mathbf{50\%}}$-$E_p$ relations (Fig. 3(b)). The overall trend is summarized in Fig. 3(c), which shows the required $E_p$ to reach $\boldsymbol{I}_{\mathbf{b}}^{\mathbf{50\%}}$ = 2.60 μA as a function of $t_p$. The increase of $E_p$ with $t_p$ reflects the finite temporal width of the excitation, which introduces a competition between energy deposition and thermal relaxation. In the limit of $t_p \rightarrow 0$, energy loss during pulse absorption becomes negligible, and $E_p$ is given by $\int_{T_0}^{T_{peak}} \boldsymbol{C_e(T)dT}$. For a given $\boldsymbol{I}_{\mathbf{b}}^{\mathbf{50\%}}$, $T_{peak}$ required for detection is fixed. Accordingly, $C_e$ determines the intercept of the $E_p$-$t_p$ relation. For finite $t_p$, energy loss through electron–phonon cooling during the pulse being absorbed reduces $T_{peak}$, requiring a larger $E_p$ to reach the same switching condition. This leads to an approximately linear increase of $E_p$ with $t_p$ with a slope governed by $\Sigma$. The simulation (solid line in Fig. 3(c)), with $\alpha$ = 8.9 and $\Sigma$ = 1.03 W m$^{-2}$ K$^{-3}$, reproduces the experimental data (see SI figure S5 for simulations at other $\boldsymbol{I}_{\mathbf{b}}^{\mathbf{50\%}}$ values). For $t_p$ = 2 ns, the required $E_p$ differs by only ~20 % from the extrapolated $t_p \rightarrow 0$ limit, indicating that the deviation from the ideal case is modest. At shorter pulse widths ($t_p$ < 2 ns), a saturation-like behaviour is observed, likely originating from distortions in the microwave delivery chain rather than the intrinsic detector response. These results show that the pulse-based calibration provides a controlled approximation of single energy-deposition events with a well-defined systematic uncertainty.

## Thermal relaxation and parameter extraction

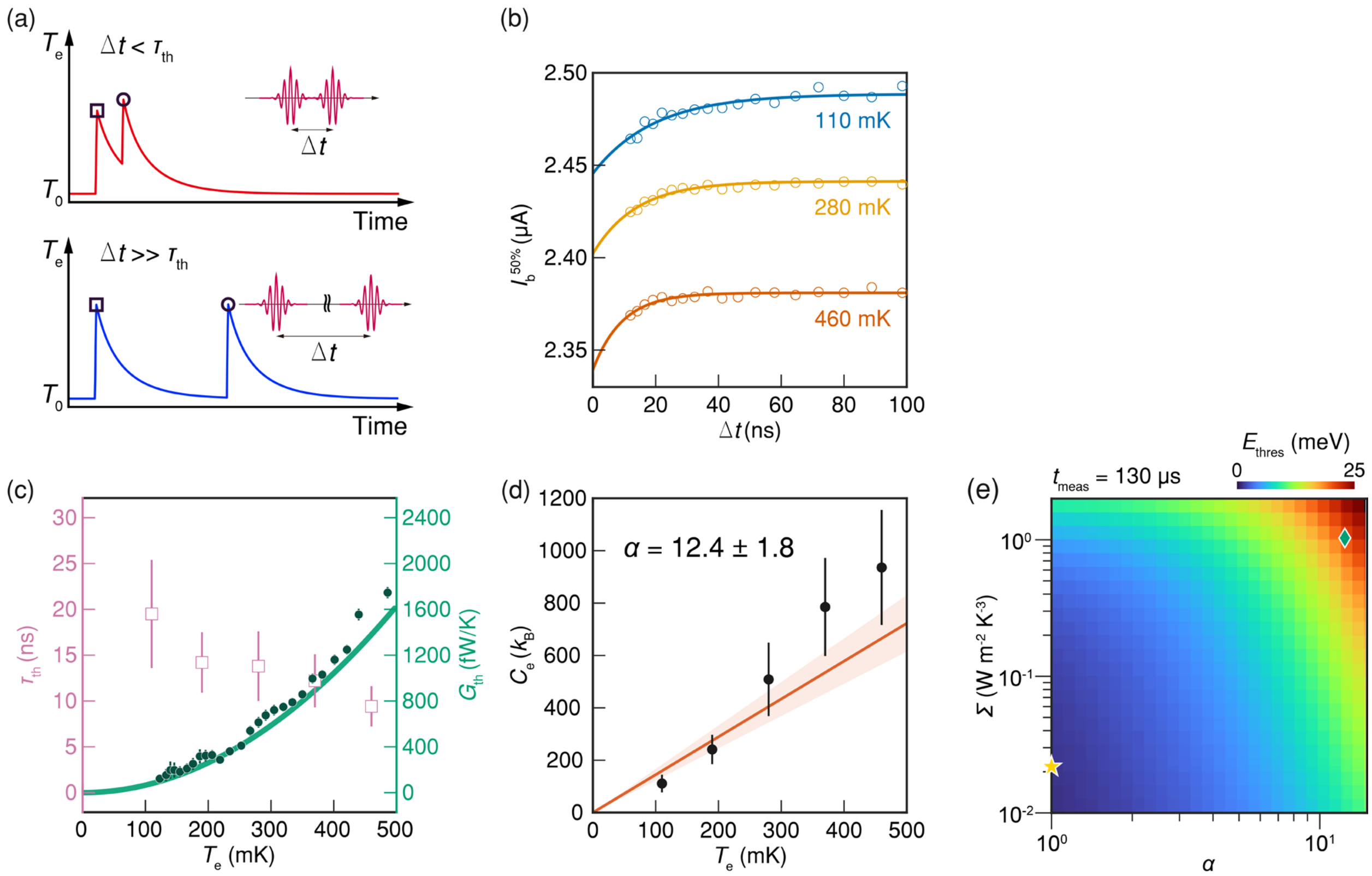

**Figure 4.** Thermal parameter extraction and its impact on the detection threshold. (a) Schematic of the electron temperature ($T_e$) dynamics following two identical pulses separated by a delay ($\Delta t$). The square and circle mark the peak $T_e$ after the first and second pulses, respectively. Upper and lower panels correspond to $\Delta t < \tau_{th}$ and $\Delta t \gg \tau_{th}$. (b) Bias current for $\eta = 0.5$ ($I_b^{50\%}$) as a function of $\Delta t$ at fixed $t_p = 1.4$ ns and $E_p = 26.8$ meV for various $T_e$. The solid line is an inverse exponential fit to the data for $\Delta t \geq 12$ ns to extract $\tau_{th}$. (c) $\tau_{th}$ (left axis) and thermal conductance ($G_{th}$, right axis) as functions of $T_e$. The error bars on $\tau_{th}$ represent 95 % confidence intervals from the single-exponential fits. The error bars on $G_{th}$ are obtained by propagating the uncertainty in $T_e$ through the numerical differentiation of the input power, which is taken to be exact; the uncertainty in $T_e$ itself originates from the standard error of the mean of the switching current averaged over 1,378 measurements. (d) Electronic heat capacity ($C_e$) as a function of $T_e$. Symbols and error bars show

experimentally extracted values and their uncertainties (95 % confidence intervals). The orange line represents the theoretically expected $C_e$ multiplied by the fitted scaling factor ($\alpha$), with the shaded region indicating 95 % confidence interval of the fit. (e) Simulated energy threshold ($E_{thres}$) as a function of multiplicative scaling factor ($\alpha$) and electron-phonon coupling parameter ($\Sigma$) for a measurement integration time ($t_{meas}$) of 130 μs. In panel (e), the yellow star marks the theoretical parameters ($\alpha$ = 1, $\Sigma$ = 0.021 W $m^{-2}K^{-3}$) and the green diamond marks the experimentally determined values ($\alpha$ = 12.4, $\Sigma$ = 1.03 W $m^{-2}K^{-3}$).

To quantitatively extract $\tau_{th}$, we employ a time-resolved two-pulse measurement[9, 11, 13, 25] (Fig. 4(a)). Two identical pulses are applied with a controllable delay ($\Delta t$). When $\Delta t < \tau_{th}$, the electron system has not fully relaxed before the second pulse arrives, and residual heat lowers the effective switching threshold, resulting in a reduced $I_b^{50\%}$. In contrast, for $\Delta t \gg \tau_{th}$, $T_e$ returns to $T_0$ before the second pulse, and the detector response approaches the single-pulse limit. Figure 4(b) shows $I_b^{50\%}$ as a function of $\Delta t$ for various $T_e$. As $\Delta t$ increases, $I_b^{50\%}$ gradually approaches the single-pulse limit ($I_{b0}^{50\%}$), indicating that the two pulses become independent after thermal relaxation. This approach follows a single exponential, $I_b^{50\%}(\Delta t) = I_{b0}^{50\%} - Ae^{-\Delta t/\tau_{th}}$, where $A$ is a free fit parameter, from which we extract $\tau_{th}$ = 19.5 (±5.9) ns (95% confidence interval) at $T_0$ = 110 mK. We restrict the fit to $\Delta t \geq 12$ ns, where the simulated temperature rise is about 150 mK, comparable to $T_0$ = 110 mK, so that the electron–phonon cooling dynamics can be approximately linearized, and a single relaxation time is well defined. Applying the same analysis at 280 mK and 460 mK yields progressively shorter $\tau_{th}$, consistent with enhanced electron–phonon cooling at higher $T_e$. Figure 4(c) shows the $T_e$ dependence of $G_{th}$ and $\tau_{th}$. We obtain $G_{th}$ from independent bolometric measurements by differentiating the input power with respect to $T_e$. The input power is well described by $P = \Sigma A(T_e^3 - T_0^3)$, giving $G_{th} = 3\Sigma A T_e^2$ (solid curve in Fig. 4(c)). From this fit we extract an electron–phonon coupling constant $\Sigma$ = 1.03 (±0.02) W m$^{-2}$ K$^{-3}$, roughly 50 times the theoretically predicted value[20] of 0.021 W m$^{-2}$ K$^{-3}$. Consistently, $\tau_{th}$ decreases progressively with increasing $T_e$, reflecting the enhancement of electron–phonon cooling through the relation $\tau_{th} = C_e/G_{th}$. Combining this $\tau_{th}$ with $G_{th} = 3\Sigma A T_0^2$, we estimate $C_e = G_{th}\tau_{th}$ over the temperature range 110-460 mK (Fig. 4(d)). $C_e$ increases linearly with $T_e$, following the theoretically expected form for graphene but exceeding it in magnitude by a factor of $\alpha$ = 12.4 (±1.8). This enhancement is corroborated by the $E_p$–$t_p$ fitting in Fig. 3(c), which independently yields $\alpha$ = 8.9 and $\Sigma$=1.03 W m$^{-2}$ K$^{-3}$; both analyses indicate an order-of-magnitude enhancement of $C_e$ compared to a theoretical expectation, demonstrating the internal consistency of the calibration framework.

The experimentally extracted parameters significantly deviate from theoretical expectations. The enhanced $C_e$ reduces the electron-temperature rise, while the measured $\Sigma$ is approximately ~50 times larger than the theoretical value, indicating stronger electron–phonon coupling and faster energy relaxation. Applying the same $E_{thres}$ criterion as in the experiment ($P_{dark}$ < 5 %), Thermal simulation using theoretical parameters predicts an ideal $E_{thres}$ of 1.28 meV (yellow star in Fig. 4e). In contrast, when the experimentally extracted $\alpha$ and $\Sigma$ are used, the simulation yields $E_{thres}$ = 21.2 meV (green diamond in Fig. 4e), in good agreement with the directly measured value of 18.1 meV. This agreement shows that the enhanced thermal parameters quantitatively account for the discrepancy between the ideal theoretical prediction and the experimentally observed sensitivity. The origin of the enhanced $C_e$ and $\Sigma$ may be associated with disorder-related effects, such as edge disorder or impurities, as suggested by previous studies in graphene[30-32] and metallic nanostructures[33]. We also note that the experimentally extracted $C_e$, $\Sigma$, and $E_{thres}$ likely represent upper bounds of the intrinsic parameters because our analysis assumes that the absorbed pulse energy is fully deposited into the graphene electron system. If a finite fraction of the incident energy is dissipated through additional channels, the actual thermal parameters would be correspondingly smaller. In addition, contact-induced doping in the graphene channel may further contribute to an overestimation of $C_e$[34-36]. Nevertheless, even under these conservative assumptions, the observation that $C_e$ and $\Sigma$ exceed naïve theoretical expectations by more than an order of magnitude highlights the need for systematic investigation of quantum limited thermal processes in graphene, including the roles of electron-electron interactions, electron-phonon coupling and disorder.

**Conclusions**

In conclusion, we demonstrated a GJJ-based calorimetric detector and established a direct energy-calibration framework using controlled microwave pulses. This approach enables self-consistent extraction of the detection threshold and key thermal parameters within a single device, while providing direct access to electronic heat capacity and thermal conductance at millikelvin temperatures, where both quantities are inherently small and difficult to measure. We obtain an energy threshold of $E_{thres}$ = 18.1 meV and a thermal relaxation time $\tau_{th}$ = 19.5 ns, combining meV-scale sensitivity with nanosecond-scale recovery, which places our device at the forefront of state-of-the-art calorimetric detectors [2, 5-7, 15, 18, 19].

Beyond demonstration of meV energy sensitivity, our results reveal the physical origin of the limited sensitivity in graphene calorimeters. Both the electronic heat capacity and the electron–phonon coupling of graphene are found to be substantially larger than theoretical expectations [20]. The enhanced $C_e$ suppresses the temperature rise for a given absorbed energy, while the elevated $\Sigma$ accelerates the energy relaxation, together reducing the calorimetric response. These effects provide a quantitative explanation for the breakdown of the ideal calorimetric scaling and the long-standing discrepancy between predicted [12] and experimentally realized sensitivities. Similar anomalous thermal dissipation has been reported for multiple graphene-based bolometric platforms in the literature [14, 16, 31, 37, 38], suggesting that such behavior is not device-specific but reflects broader characteristics of the graphene platform.

More broadly, graphene Josephson junctions provide a compact microwave calorimeter architecture that combines broadband microwave absorption, superconducting threshold readout, and nanosecond thermal recovery. Future improvements will require not only optimizing the Josephson switching threshold but also engineering the microscopic thermal environment of graphene, including contacts, edges, disorder, and carrier-density profile. These results establish graphene Josephson junctions as a versatile platform for microwave calorimetry and provide design principles for next-generation quantum sensors based on low-dimensional materials.

## Methods

### Sample Fabrication

The hBN, mono-layer graphene (MLG), and graphite flakes were exfoliated on a Si/SiO2 wafer. All stacks used in this experiment were dry-transferred using polymer (Elvacite 2552C) stamps and annealed in vacuum at 500 °C for 2 h to remove bubbles and surface impurities. Clean areas free of residues and bubbles were identified using an atomic force microscope (AFM, XE7 from Park Systems) topography imaging. The thickness of hBN used for the stack was measured by AFM. We prepared the input resonators and the DC line with a mask aligner (MA8, Suss Microtec photo-lithography) on an undoped/intrinsic Si wafer (525 μm) whose backside was coated with Cr (5 nm) and Au (100 nm) for ground. For all electrodes, including the input resonator, DC lines and bottom gate line, MoRe (100 nm) was sputtered as the superconducting material. After the process, the stack was dry-transferred between the input resonators. Electrodes and etching masks were patterned by standard electron-beam lithography. The hBN was selectively etched under CF4 plasma, and the MLG and graphite under O2 plasma, using a reactive ion etcher (RIE). Electrical isolation between the bottom graphite gate and the MLG was achieved with overdosed cross-linked poly(methyl methacrylate) (PMMA). MoRe (75 nm) was sputtered for the contacts of the Josephson junction, input resonators and bottom gate. For bandage interconnects between photolithography- and electron-beam-lithography-defined electrodes, PMMA was patterned and developed to open windows over the bandage areas. The exposed surfaces were Ar-ion-milled to remove native oxides, and an additional MoRe layer (170 nm) was sputtered to complete the electrical bridge.

**Transport Measurement**

The microwave input resonator coupled to the graphene was characterized using a vector network analyzer (VNA, E5063A, Keysight). The bottom-gate voltage was applied using a voltage source (Yokogawa 7651). The graphene Josephson junction (GJJ) was current-biased using an arbitrary waveform generator (33512B, Keysight), which also provided the trigger for a second AWG (AWG7102, Tektronix) used to generate the excitation pulses. The junction voltage was amplified with a voltage amplifier (SP1004, Basel Precision Instruments) and measured using a digital multimeter (Keithley 2000).

**Pulse Generation**

Two Gaussian pulses with a center frequency of 1 GHz and a relative phase difference of 90° were applied to the I and Q ports of an IQ mixer (MMIQ-0416LS, Marki Microwave). By mixing these pulses with a continuous 8.64 GHz microwave signal from a signal generator (E8257D, Keysight), we generated a Gaussian pulse with a center frequency of 7.64 GHz.

**Supporting Information.**

Input-resonator reflectometry, determination of the base electron temperature, bolometric measurements, and numerical modeling of the detection threshold (PDF).

**Author Information**

Corresponding Author

Correspondence to Gil-Ho Lee (lghman@postech.ac.kr)

**Author Contributions**

G.-H.L. and S.L. conceived the study. S.L. fabricated the devices with assistance from Seongeun J. and Seong J. S.L. performed the measurements. G.-H.L., S.L., J.K., J.-C.P., and P.K. analyzed the data. S.L., W.J., J.K., and D.T.P. built the low-noise measurement setup. D.T.P. measured the electron temperature of the measurement setup. K.C.F. provided expertise in RF techniques. K.W. and T.T. grew hBN crystals. G.-H.L. supervised the work. G.-H.L. and S.L. wrote the manuscript. All authors reviewed and approved the manuscript.

**Acknowledgments**

We acknowledge helpful discussions with Junho Suh on RF measurement, Jinho Park for valuable discussions on graphene Josephson junctions and pulse generation, and Yunchul Chung for providing quantum dot devices used for electron temperature calibration. This research was supported by the Samsung Science and Technology Foundation (Project No. SSTF-BA2101-06) and the Agency for Defense Development through a grant funded by the Defense Acquisition Program Administration (DAPA) (UI257011TE). G.-H.L., S.L., and W.J. acknowledge the support by the National Research Foundation of Korea (NRF) (nos., RS-2022-NR068223, RS-2024-00393599, RS-2024-00442710, RS-2025-02317602) and by the ITRC (Information Technology Research Center) support program (IITP-2026-RS-2022-00164799). D.T.P. was partially supported by the NRF (RS-2024-00463743). Nanofabrication was partially supported by POSTECH Center for Semiconductor Technology Convergence supported by Korea Basic Science Institute, NFEC-2025-03-304835. K.W. and T.T. acknowledge support from the JSPS KAKENHI (Grant Numbers 21H05233 and 23H02052) and World Premier International Research Center Initiative (WPI), MEXT, Japan. J.-C.P. acknowledges the support by the NRF grants funded by the Ministry of Science and ICT (RS-2024-00356960) and by the Ministry of Education (RS-2025-25442707). K.C.F. acknowledges support from the Air Force Office of Scientific Research under Cooperative Agreement FA2386-25-1-4067. P. K. was supported by the Air Force Office of Scientific Research under award number FA2386-24-1-4073. The views and conclusions contained in this document are those of the authors and should not be interpreted as representing the official policies, either expressed or implied, of the Asian Office of Aerospace Research and Development or the U.S. Government.

**References**

(1) Gol'Tsman, G. N.; Okunev, O.; Chulkova, G.; Lipatov, A.; Semenov, A.; Smirnov, K.; Voronov, B.; Dzardanov, A.; Williams, C.; Sobolewski, R. Picosecond superconducting single-

photon optical detector. *Applied Physics Letters* **2001**, *79* (6), 705–707. DOI: 10.1063/1.1388868.
(2) Taylor, G. G.; Walter, A. B.; Korzh, B.; Bumble, B.; Patel, S. R.; Allmaras, J. P.; Beyer, A. D.; O'Brient, R.; Shaw, M. D.; Wollman, E. E. Low-noise single-photon counting superconducting nanowire detectors at infrared wavelengths up to 29 μm. *Optica* **2023**, *10* (12), 1672–1678.
(3) Day, P. K.; Cothard, N. F.; Albert, C.; Foote, L.; Kane, E.; Eom, B. H.; Basu Thakur, R.; Janssen, R. M.; Beyer, A.; Echternach, P. M. A 25-micrometer single-photon-sensitive kinetic inductance detector. *Physical Review X* **2024**, *14* (4), 041005.
(4) Day, P. K.; Leduc, H. G.; Mazin, B. A.; Vayonakis, A.; Zmuidzinas, J. A broadband superconducting detector suitable for use in large arrays. *Nature* **2003**, *425* (6960), 817–821. DOI: 10.1038/nature02037.
(5) Gunyhó, A. M.; Kohvakka, K.; Chen, Q.-M.; Girard, J.-P.; Kokkoniemi, R.; Liu, W.; Möttönen, M. Zeptojoule calorimetry. *Nature Electronics* **2026**, *9* (6), 611–620. DOI: 10.1038/s41928-026-01615-2.
(6) Govenius, J.; Lake, R.; Tan, K.; Möttönen, M. Detection of zeptojoule microwave pulses using electrothermal feedback in proximity-induced Josephson junctions. *Physical Review Letters* **2016**, *117* (3), 030802.
(7) Echternach, P.; Pepper, B.; Reck, T.; Bradford, C. Single photon detection of 1.5 THz radiation with the quantum capacitance detector. *Nature Astronomy* **2018**, *2* (1), 90–97.
(8) Echternach, P. M.; Beyer, A. D.; Bradford, C. M. Large array of low-frequency readout quantum capacitance detectors. *Journal of Astronomical Telescopes, Instruments, and Systems* **2021**, *7* (1), 011003–011003.
(9) Yan, J.; Kim, M. H.; Elle, J. A.; Sushkov, A. B.; Jenkins, G. S.; Milchberg, H. M.; Fuhrer, M. S.; Drew, H. Dual-gated bilayer graphene hot-electron bolometer. *Nature Nanotechnology* **2012**, *7* (7), 472–478.
(10) Fong, K. C.; Schwab, K. Ultrasensitive and Wide-Bandwidth Thermal Measurements of Graphene at Low Temperatures. *Physical Review X* **2012**, *2* (3), 031006.
(11) Cai, X.; Sushkov, A. B.; Suess, R. J.; Jadidi, M. M.; Jenkins, G. S.; Nyakiti, L. O.; Myers-Ward, R. L.; Li, S.; Yan, J.; Gaskill, D. K.; et al. Sensitive room-temperature terahertz detection via the photothermoelectric effect in graphene. *Nature Nanotechnology* **2014**, *9* (10), 814–819. DOI: 10.1038/nnano.2014.182.
(12) Walsh, E. D.; Efetov, D. K.; Lee, G.-H.; Heuck, M.; Crossno, J.; Ohki, T. A.; Kim, P.; Englund, D.; Fong, K. C. Graphene-Based Josephson-Junction Single-Photon Detector. *Physical Review Applied* **2017**, *8* (2), 024022. DOI: 10.1103/physrevapplied.8.024022.
(13) Efetov, D. K.; Shiue, R.-J.; Gao, Y.; Skinner, B.; Walsh, E. D.; Choi, H.; Zheng, J.; Tan, C.; Grosso, G.; Peng, C.; et al. Fast thermal relaxation in cavity-coupled graphene bolometers with a Johnson noise read-out. *Nature Nanotechnology* **2018**, *13* (9), 797–801. DOI: 10.1038/s41565-018-0169-0.
(14) Lee, G.-H.; Efetov, D. K.; Jung, W.; Ranzani, L.; Walsh, E. D.; Ohki, T. A.; Taniguchi, T.; Watanabe, K.; Kim, P.; Englund, D.; et al. Graphene-based Josephson junction microwave bolometer. *Nature* **2020**, *586* (7827), 42–46. DOI: 10.1038/s41586-020-2752-4.
(15) Aamir, M. A.; Moore, J. N.; Lu, X.; Seifert, P.; Englund, D.; Fong, K. C.; Efetov, D. K. Ultrasensitive calorimetric measurements of the electronic heat capacity of graphene. *Nano Letters* **2021**, *21* (12), 5330–5337.

(16) Katti, R.; Arora, H. S.; Saira, O.-P.; Watanabe, K.; Taniguchi, T.; Schwab, K. C.; Roukes, M. L.; Nadj-Perge, S. Hot Carrier Thermalization and Josephson Inductance Thermometry in a Graphene-Based Microwave Circuit. *Nano Letters* **2023**, *23* (10), 4136–4141.
(17) Kokkoniemi, R.; Girard, J.-P.; Hazra, D.; Laitinen, A.; Govenius, J.; Lake, R.; Sallinen, I.; Vesterinen, V.; Partanen, M.; Tan, J. Bolometer operating at the threshold for circuit quantum electrodynamics. *Nature* **2020**, *586* (7827), 47–51.
(18) Walsh, E. D.; Jung, W.; Lee, G.-H.; Efetov, D. K.; Wu, B.-I.; Huang, K. F.; Ohki, T. A.; Taniguchi, T.; Watanabe, K.; Kim, P.; et al. Josephson junction infrared single-photon detector. *Science* **2021**, *372* (6540), 409–412. DOI: 10.1126/science.abf5539.
(19) Huang, B.; Arnault, E. G.; Jung, W.; Fried, C.; Russell, B. J.; Watanabe, K.; Taniguchi, T.; Henriksen, E. A.; Englund, D.; Lee, G.-H.; et al. Thermal detection of single photons using Dirac fermions. *Nature Communications* **2026**, *17* (1), 3845. DOI: 10.1038/s41467-026-70648-0.
(20) Viljas, J.; Heikkilä, T. Electron-phonon heat transfer in monolayer and bilayer graphene. *Physical Review B* **2010**, *81* (24), 245404.
(21) Song, J. C.; Reizer, M. Y.; Levitov, L. S. Disorder-assisted electron-phonon scattering and cooling pathways in graphene. *Physical Review Letters* **2012**, *109* (10), 106602.
(22) Chen, W.; Clerk, A. A. Electron-phonon mediated heat flow in disordered graphene. *Physical Review B* **2012**, *86* (12), 125443.
(23) Borzenets, I. V.; Coskun, U. C.; Mebrahtu, H. T.; Bomze, Y. V.; Smirnov, A. I.; Finkelstein, G. Phonon Bottleneck in Graphene-Based Josephson Junctions at Millikelvin Temperatures. *Physical Review Letters* **2013**, *111* (2), 027001. DOI: 10.1103/physrevlett.111.027001.
(24) Betz, A. C.; Jhang, S. H.; Pallecchi, E.; Ferreira, R.; Fève, G.; Berroir, J. M.; Plaçais, B. Supercollision cooling in undoped graphene. *Nature Physics* **2013**, *9* (2), 109–112. DOI: 10.1038/nphys2494.
(25) Graham, M. W.; Shi, S.-F.; Ralph, D. C.; Park, J.; McEuen, P. L. Photocurrent measurements of supercollision cooling in graphene. *Nature Physics* **2013**, *9* (2), 103–108. DOI: 10.1038/nphys2493.
(26) Courtois, H.; Meschke, M.; Peltonen, J. T.; Pekola, J. P. Origin of Hysteresis in a Proximity Josephson Junction. *Physical Review Letters* **2008**, *101* (6), 067002. DOI: 10.1103/physrevlett.101.067002.
(27) Voutilainen, J.; Fay, A.; Häkkinen, P.; Viljas, J. K.; Heikkilä, T. T.; Hakonen, P. J. Energy relaxation in graphene and its measurement with supercurrent. *Physical Review B* **2011**, *84* (4), 045419.
(28) Koechlin, F.; Bonin, B. Parametrization of the niobium thermal conductivity in the superconducting state. *Superconductor Science and Technology* **1996**, *9* (6), 453–460. DOI: 10.1088/0953-2048/9/6/003.
(29) Zgirski, M.; Foltyn, M.; Savin, A.; Norowski, K.; Meschke, M.; Pekola, J. Nanosecond thermometry with Josephson junctions. *Physical Review Applied* **2018**, *10* (4), 044068.
(30) Allen, M. T.; Shtanko, O.; Fulga, I. C.; Akhmerov, A.; Watanabe, K.; Taniguchi, T.; Jarillo-Herrero, P.; Levitov, L. S.; Yacoby, A. Spatially resolved edge currents and guided-wave electronic states in graphene. *Nature Physics* **2016**, *12* (2), 128–133.
(31) Halbertal, D.; Ben Shalom, M.; Uri, A.; Bagani, K.; Meltzer, A. Y.; Marcus, I.; Myasoedov, Y.; Birkbeck, J.; Levitov, L. S.; Geim, A. K.; et al. Imaging resonant dissipation from individual atomic defects in graphene. *Science* **2017**, *358* (6368), 1303–1306. DOI: 10.1126/science.aan0877.

(32) Jeong, H.-W.; Jang, S.; Park, S.; Watanabe, K.; Taniguchi, T.; Lee, G.-H. Edge dependence of nonlocal transport in gapped bilayer graphene. *Nano Letters* **2024**, *24* (50), 15950–15955.
(33) Viisanen, K.; Pekola, J. Anomalous electronic heat capacity of copper nanowires at sub-kelvin temperatures. *Physical Review B* **2018**, *97* (11), 115422.
(34) Allen, M. T.; Shtanko, O.; Fulga, I. C.; Wang, J. I. J.; Nurgaliev, D.; Watanabe, K.; Taniguchi, T.; Akhmerov, A. R.; Jarillo-Herrero, P.; Levitov, L. S.; et al. Observation of Electron Coherence and Fabry–Perot Standing Waves at a Graphene Edge. *Nano Letters* **2017**, *17* (12), 7380–7386. DOI: 10.1021/acs.nanolett.7b03156.
(35) Park, J.; Lee, J. H.; Lee, G.-H.; Takane, Y.; Imura, K.-I.; Taniguchi, T.; Watanabe, K.; Lee, H.-J. Short Ballistic Josephson Coupling in Planar Graphene Junctions with Inhomogeneous Carrier Doping. *Physical Review Letters* **2018**, *120* (7), 077701. DOI: 10.1103/physrevlett.120.077701.
(36) Jang, S.; Park, G.-H.; Park, S.; Jeong, H.-W.; Watanabe, K.; Taniguchi, T.; Lee, G.-H. Engineering Superconducting Contacts Transparent to a Bipolar Graphene. *Nano Letters* **2024**, *24* (49), 15582–15587. DOI: 10.1021/acs.nanolett.4c03767.
(37) Kong, J. F.; Levitov, L.; Halbertal, D.; Zeldov, E. Resonant electron-lattice cooling in graphene. *Physical Review B* **2018**, *97* (24), 245416.
(38) Draelos, A. W.; Wei, M.-T.; Seredinski, A.; Li, H.; Mehta, Y.; Watanabe, K.; Taniguchi, T.; Borzenets, I. V.; Amet, F.; Finkelstein, G. Supercurrent flow in multiterminal graphene Josephson junctions. *Nano Letters* **2019**, *19* (2), 1039–1043.